\documentclass[12pt]{iopart} 
\usepackage{float}
\usepackage{iopams}  
\usepackage{graphicx}
\begin{document}
\title{Acceleration of macroscopic clusters in crossed magnetic fields}
\author{A.R. Karimov$^{1,2}$, S.A. Terekhov$^{2^*}$, A.E. Shikanov$^{2}$, P.A. Murad$^{3}$}
\address{$^1$Institute for High Temperatures RAS,  Izhorskaya 13/19, Moscow 127412, Russia \\
$^2$National Research Nuclear University MEPhI, Kashirskoye shosse 31, Moscow, 115409, Russia\\
$^3$ Morningstar Applied Physics, LLC Vienna, VA 22182 USA\\
$^*$email: saterekhov@mephi.ru}
\begin{abstract}
In previous our investigations \cite{Karimov, KM}, the acceleration of rotating plasma flow in crossed magnetic fields owing to the momentum transfer between the macroscopic degrees of freedom for the plasma flow was studied. Based on these results, we have discussed the acceleration of plasma flow containing the charged macroscopic particles. The estimates presented make it possible to draw a tentative conclusion about the axial acceleration of the dusty plasma flows.
\end{abstract}


\section{Introduction}

The acceleration of rotating plasma flows consisting of electrons, ions and macroscopic bodies in crossed electric and magnetic fields is of interest for different technical applications, including plasma thrusters \cite{Karimov, KM}. Method of acceleration of macroscopic bodies in rail electrodynamic accelerators based on the transformation of the electrodynamic energy of the current pulse into kinetic energy due to the action of the Ampere force on the plasma bunch has the advantage over other methods of creating high-speed matter flows required for various applications \cite{Arcimovich}-\cite{Galanin}. In this case, the magnitude of the achievable rate of throwing macro bodies is determined only by the processes of accumulation and release of electrical energy: to achieve high speeds, it is necessary to increase the power of the pulsed current sources used, which has technical limitations \cite{Nosov}. Another cause limiting the maximum speed is the developing instability of the plasma piston, which is most strongly manifested when the current in the accelerator circuit changes \cite{Bobashev}. 

These features are absent in quasistationary plasma-optical systems, e.g. plasma thrusters. However, in contrast to the rail electrodynamic accelerator, where a powerful current pulse creates a plasma piston pushing the macroscopic body \cite{Kaznelson}-\cite{Plekhanov}, here we should create conditions when macroscopic particles will be entrained by the plasma flow.

In this paper, we are going to consider this possibility in the scheme of a plasma accelerator using the acceleration of a rotating plasma flow in crossed magnetic fields due to momentum transfer between the macroscopic degrees of freedom of the plasma flow, which in essence the same as the mechanism in rail electrodynamic accelerators.

\section{Schematic of plasma accelerator}

In this paper, we will study the opposite limit when the displacement current is compensated by the conduction current. Such situation may be realized if we can neglect by the space dependence of a magnetic field. It means that the spatial scales of the acceleration domain are large enough to escape the influence of the border effects while the temporal dependence of the  magnetic self-field is completely defined by the azimuthal component of the electric self-field.

\begin{figure}[H]
\begin{minipage}[h]{0.47\linewidth}
\begin{center}
\includegraphics[width=8cm]{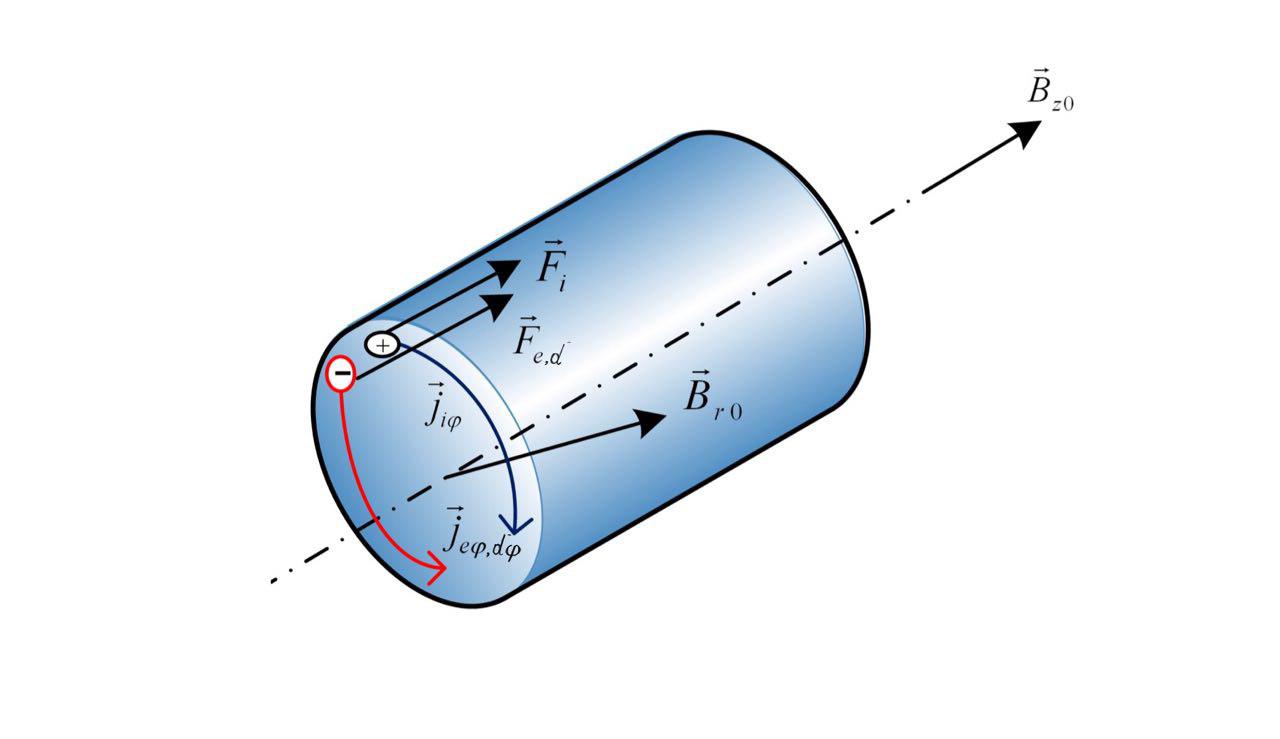} \\(a)
\end{center}
\end{minipage}
\hfill
\begin{minipage}[h]{0.47\linewidth}
\begin{center}
\includegraphics[width=8cm]{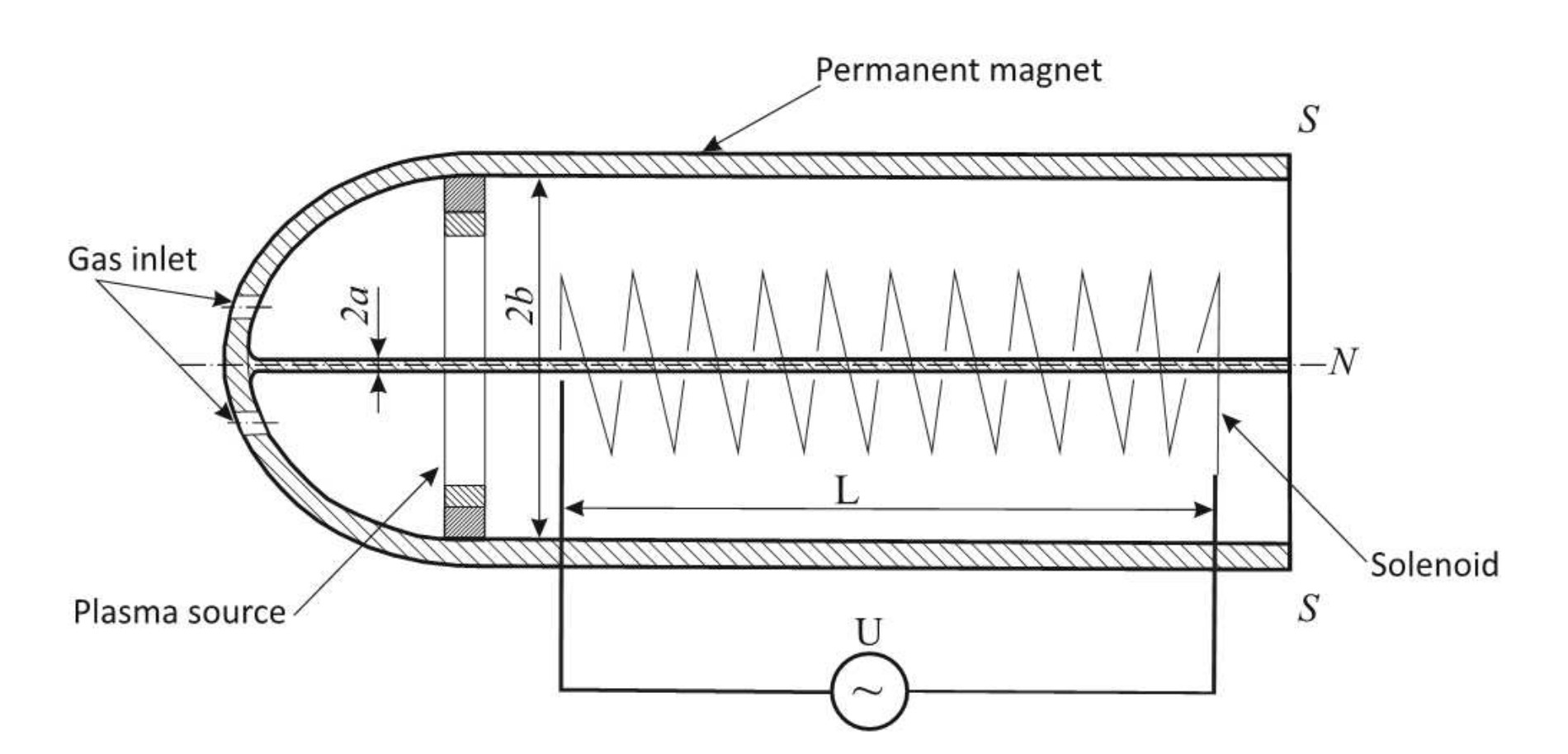} \\(b)
\end{center}
\end{minipage}
\caption{Schematic of electron, ion and macroscopic bodies acceleration region (a). Accelerator design (b). \label{accelerator}}
\end{figure}

Let us investigate this possibility for an already formed collisionless plasma flow containing electrons, positive singly charged ions and heavy multiply charged negative particles, applied to an accelerator based on nonequilibrium transfer of momentum between the macroscopic degrees of freedom of the plasma flow in crossed magnetic fields of a special type (see Fig. \ref{accelerator}(a))\cite{Karimov, KM}. One can make such a configuration of magnetic fields with the help of a solenoid ($B_{z0}$) and permanent magnets ($B_{r0}$). The change of axial magnetic flux induces an azimuthal electric field that rotates electron ${\bf j}_{e \varphi}$, dust particle ${\bf j}_{d \varphi}$ and ion ${\bf j}_{i \varphi}$ flows in different directions as seen in Fig. \ref{accelerator}(a). The azimuthal electron and ion flows interact with the radial component $B_{r0}$ via the Lorentz force, this leads to the axial acceleration of electrons, dust particles and ions in the same axial direction since the resulting forces ${\bf j}_{e \varphi}\times {\bf B}_{r0}$ for electron, dust particle ${\bf j}_{d \varphi}\times {\bf B}_{r0}$ and ion ${\bf j}_{i \varphi}\times {\bf B}_{r0}$ are directed in the same direction. Owing to this process, the transformation of azimuthal momentum transfers into axial momentum. Thus, the electron, dust particle and ion flows are accelerated in one axial direction.

This acceleration mechanism for the rotating plasma flow in the external magnetic fields can be strengthened in the schematic of a thruster concept sketched in Fig. \ref{accelerator}(b) \cite{KM}. Here, a gas fuel is injected into the system through the gas inlet. Then, the gas flux enters the zone of plasma source where the gas is ionized to create a plasma flux that would be used to generate thrust. In the present schematic, we have only drawn the ionization zone without discussing the physical mechanism and construction of the plasma source since it requires a separate discussion and we hope to return to the issue in our forthcoming studies.

Thus, it can be noted that the plasma acquires an azimuthal momentum from the field, and the radial magnetic field transfers it to the axial direction. In technical realization and physical essence the considered mechanism of acceleration is close to the mechanism incorporated in the rail eletrodynamic accelerator. Expected that this configuration of the external magnetic field should enhance the momentum/energy exchange due to a local quasi-neutrality violation, which leads to the appearance of a local time-dependent  electric field, which provide the redistribution of momentum/energy in space and time. 

\section{Model of plasma acceleration}

In order to understand the behavior of the flow depicted in Fig. \ref{accelerator}, we will study the following model that considers an azimuthally symmetric plasma where the derivative vanishes ($\partial_{\varphi} = 0$) but will include the azimuthal components in the electric and magnetic fields. We propose in the system there is an external spatially homogeneous magnetic field which depends only on time:
$$\mathbf{B}^0= B_{r0}(t)\mathbf{e}_r + B_{z0}(t)\mathbf{e}_z\/,$$
where $B_{r0}(t)$ and $B_{z0}(t)$ are some known functions. Such an  external magnetic field induces an electric field which can  possibly rotate electrons and ions in different directions if the model is successful. 

In view of linearity of the Maxwell equations, we can present the total electric and magnetic fields of the system. Let us assume we have mean values with a perturbation such as:
$$\mathbf{E}= \mathbf{E}^0 + \mathbf{E}^{*}$$
and
$$\mathbf{B}= \mathbf{B}^0 + \mathbf{B}^{*}\/,$$
where the perturbed electric $\mathbf{E}^{*}$ and magnetic $\mathbf{B}^{*}$ fields are defined by the dynamic processes in itself plasma medium described by the standard cold-fluid model in the dimensionless form:
\begin{equation}
{\partial_{t} n_{s}}+\nabla(n_{s}{\bf v}_{s})=0,
\label{dtn}
\end{equation}
\begin{equation}
{\partial_{t} {\bf v}_{s}} + {\bf v}_{s} \nabla {\bf v}_{s}=\mu_{s} \left({\bf E} + \left[{\bf v}_{s} \times {\bf B} \right] \right),
\label{dtv}
\end{equation}
where the label $s$ relates to the type of paticles: $e$ relates to the electron fluid, $i$ - to the ion fluid and $d$ - to the negative charged macroparticle fluid. In this case $q_{e}=-e$, $q_{i}=e$ (in the case of a single-charged plasma surrounding the macroparticle), and $q_{d}=-Z_{d}e$; $\mu_{e}=-1$, $\mu_{i}=m_{e}/m_{i}$ and $\mu_{d}=-Z_{d}m_{e}/m_{d}$.
The perturbed fields are defined by
\begin{equation}
{\partial_{t} {\bf E}^{*}}=n_{e}{\bf v}_{e}+Z_{d}n_{d}{\bf v}_{d}-n_{i}{\bf v}_{i},
\label{dtE*}
\end{equation}
\begin{equation}
\nabla {\bf E}^{*}=n_{i}-n_{e}-Z_{d}n_{d},
\label{divE}
\end{equation}
\begin{equation}
\nabla  {\bf B}^{*}=0,
\label{divB}
\end{equation}
\begin{equation}
\nabla \times  {\bf E}^{*}=- \frac{\partial {\bf B}^{*}}{\partial t} ,
\label{rotE}
\end{equation}
where $n_{i}$ is the ion density, $n_{e}$ is the electron density, $n_{d}$ is the dust particle density, $Z_{d}$ is the dust particle charge number, and ${\bf v}_{i}$, ${\bf v}_{e}$, ${\bf v}_{d}$ are the ion, electron and dust particle fluid velocities respectively.

In order to write these equations in terms of dimensionless variables we used the initial electron density $n_0$, the initial radius of plasma cylinder $R_0$, the inverse electron frequency $\omega_{pe}=(4\pi n_0 e^2/m_e)^{1/2}$, where $e$ is the electric charge, as the natural scale of all densities, coordinates and time. All velocities are normalized by the $R_0\omega_{pe}$, the electric field is normalized by the $4\pi e n_0 R_0$ and  the magnetic field is normalized by the $4\pi e n_0 R_0 c /(R_0\omega_{pe})$, where $c$ is the light velocity.

From equation of induction (\ref{rotE}) written for external magnetic field $\mathbf{B}^0$ in integral form 
$$\int_{\gamma} \mathbf{E}^0d\mathbf{l}= - \partial_t \left( \int_{S_{\gamma}}\mathbf{B}^0 d\mathbf{S}\right)$$
it follows that the external electric field has got only an azimuthal component which is determined as
\begin{equation}
E_{\phi_{0}}=- \frac{r}{2} \partial_{t} B_{z_{0}} .
\label{Ephi0}
\end{equation}
This relation suggests to seek the possible solution of governing set in the form
\begin{equation}
{\bf v}_{s}= r A_{s}(t){\bf e}_{r} + r C_{s}(t){\bf e}_{\phi} + r D_{s}(t){\bf e}_{z},
\label{v_s}
\end{equation}
where $A_s(t)$, $C_s(t)$ and $D_s(t)$ are associated with the radial, axial, and azimuthal velocity components and are still to be determined. Clearly, the simple basis structure proposed in Eq. (\ref{v_s}) is not unique, however, depending on the initial data $A_s(t=0)$, $C_s(t=0)$ and $D_s(t=0)$, this relation describes a rich variety of temporal dependences.

Also, supposing that the intrinsic electric field of the system is similar to the velocity field (\ref{v_s}), we can write
\begin{equation}
{\bf E}^{*}= r \varepsilon_{r}(t){\bf e}_{r} + r \varepsilon_{\phi}(t){\bf e}_{\phi} + r \varepsilon_{z}(t){\bf e}_{z},
\label{E*}
\end{equation}
and set
\begin{equation}
n_{s}=n_{s}(t).
\label{n(t)}
\end{equation}
In this case, from (\ref{rotE}) it follows that
\begin{equation}
\partial_t B_r^{*} = 0, \hspace{5mm} \partial_t B_{\varphi}^{*} = 0, \hspace{5mm} \partial_t B_z^{*} = -2 \varepsilon_{\varphi}\/,
\label{11_ac}
\end{equation}
i.e. for the initial conditions $$B_r^{*}(t=0)=B_{\varphi}^{*}(t=0)=B_z^{*}(t=0)=0$$
only the component $B_z^{*}$ may be distinct from zero inside the plasma cylinder and the one  is the function of time. It means that $\nabla \times \mathbf{B}^{*}\equiv 0$. 
Moreover, substituting (\ref{E*}) and (\ref{n(t)}) into Eq. (\ref{divE}) we get the temporal relation between the electron and ion densities 
\begin{equation}
n_{i}=n_{e}+Z_{d}n_{d}-2\varepsilon_{r}.
\label{divEcill2}
\end{equation}
It should be noted that at this point we have given up the routine quasi-neutrality condition ($n_i=n_e$) that may bring about new features of the plasma dynamics. The relation (\ref{divEcill2}) assumes the current can move radially between the anode and the cathode.  In this case, it becomes possible to generate perturbed electric fields, which can be responsible for the capture and acceleration of dusty plasma components in the accelerating structure.

Inserting (\ref{n(t)}) - (\ref{divEcill2}) into (\ref{dtn})-(\ref{rotE}) yields

\begin{equation}
\frac {dn_{s}}{dt}+2n_{s}A_{s}=0,
\label{dtn2}
\end{equation}
\begin{equation}
\frac{dA_{s}}{dt} + A^{2}_{s}-C^{2}_{s}=\mu_{s} \left[\varepsilon_{r} + C_{s}(B_{z_{0}}+B^{*}_{z}) \right],
\label{dtAs}
\end{equation}
\begin{equation}
\frac{dC_{s}}{dt} + 2A_{s}C_{s}= \mu_{s} \left[\varepsilon_{\phi} - \frac{1}{2}\frac{dB_{z_{0}}}{dt} + D_{s}B_{r_{0}} - A_{s}(B_{z_{0}}+B^{*}_{z}) \right],
\label{dtCs}
\end{equation}
\begin{equation}
\frac{dD_{s}}{dt} + A_{s}D_{s}=\mu_{s} \left[\varepsilon_{z} - C_{s}B_{r_{0}} \right],
\label{dtDs}
\end{equation}
\begin{equation}
\frac{d\varepsilon_{r}}{dt}=n_{e}(A_{e}-A_{i})+Z_{d}n_{d}(A_{d}-A_{i}) - 2\varepsilon_{r}A_{i},
\label{dtEr}
\end{equation}
\begin{equation}
\frac{d\varepsilon_{\phi}}{dt}=n_{e}(C_{e}-C_{i})+Z_{d}n_{d}(C_{d}-C_{i}) - 2\varepsilon_{r}C_{i},
\label{dtEphi}
\end{equation}
\begin{equation}
\frac{d\varepsilon_{z}}{dt}=n_{e}(D_{e}-D_{i})+Z_{d}n_{d}(D_{d}-D_{i}) - 2\varepsilon_{r}D_{i},
\label{dtEz}
\end{equation}
\begin{equation}
\nabla {\bf B}^{*}\equiv 0,
\label{divBcill2}
\end{equation}
\begin{equation}
\frac{d B^{*}_{z}}{dt}=-2\varepsilon_{\phi}.
\label{rotEcill2}
\end{equation}
This system of ordinary differential nonlinear equations (\ref{dtn2})-(\ref{rotEcill2}) is describing the dynamic of electrons, ions and negatively charged macroscopic particles in crossed magnetic fields. In previous investigations this system was used for describing of dynamic processes in two-component plasma flows \cite{Karimov}.

\section{Capture and acceleration of macroscopic bodies}

To demonstrate the possibility of capture and acceleration of macroscopic particles contained in a dusty plasma, we consider an argon plasma, consisting massive negative charging macroscopic particles with weight $m_{d}=10^{3}m_{p}$, where $m_{p}$ - the proton mass \cite{FHH_2004}-\cite{SuLam_1963}. We took the initial electron dencity value $n_{e0}=10^{16}$ sm$^{-3}$, and initial rations of ions and macroparticles were $n_{i0}/n_{e0}=1,01$ and $n_{d0}/n_{e0}=10^{-5}$ respectively. These rations of dencities of the dusty plasma components is typical for experiments, where the charging process of macroparticles depends on mobility of electrons and ions \cite{FHH_2004}-\cite{Shukla}. In this case an estimate of cluster size $a=6,068 \times 10^{-9}$ m and the value of its negative charge $Z_{d}=998,26 e$, obtained for a locally equilibrium plasma, coincide with experimental values \cite{FHH_2004}-\cite{Shukla}.

First in the calculations was taken into account the dependence of the axial velocity of the components of the dusty plasma on the external axial magnetic field. In these calculations the permanent radial magnetic field was directed into the plasma cylinder, which requires $B_{r0}=-1$.The initial velocitiies of electrons, ions and macroparticles were next: $A_{e}(0)=A_{i}(0)=A_{d}=-10^{-2}$, i.e. the radial velocity components were directed into the plasma cylinder, $-C_{e}(0)=-C_{d}=C_{i}(0)=10^{-3}$ and $D_{e}(0)=D_{i}(0)=D_{d}=10^{-1}$. The initial values of the perturbed electric fields were $\varepsilon_{r}(0)=0$, $\varepsilon_{\phi}(0)=0$ and $\varepsilon_{z}(0)=0$, and the initial value of the perturbed axial magnetic field was $B^{*}_{z}(0)=0$.

\begin{figure}[H]
\begin{minipage}[h]{0.47\linewidth}
\begin{center}
\includegraphics[width=7cm]{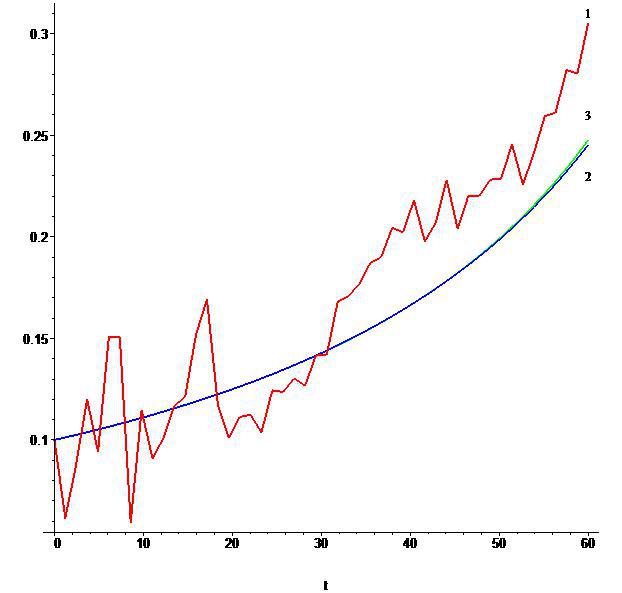} \\ (a)
\end{center}
\end{minipage}
\hfill
\begin{minipage}[h]{0.47\linewidth}
\begin{center}
\includegraphics[width=7cm]{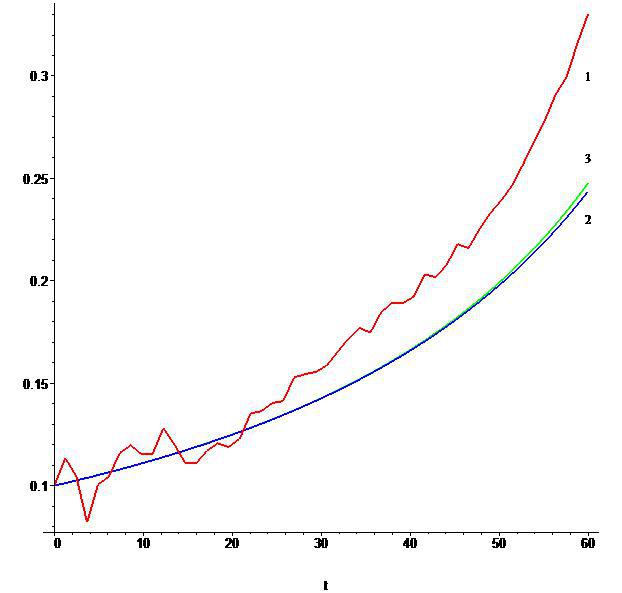} \\ (b)
\end{center}
\end{minipage}
\caption{Temporal dependence of $D_{e}$ (curve 1), $D_{i}$ (curve 2) and $D_{d}$ (curve 3) for $A_{e}(0)=A_{i}(0)=A_{d}=-10^{-2}$, $-C_{e}(0)=-C_{d}=C_{i}(0)=10^{-3}$ for the case $B_{r0}=-1$ and $B_{z0}=1$ (a) and for the case $B_{r0}=-1$ and $B_{z0}=-0.25t$ (b). \label{Pz}}
\end{figure}

In Fig. \ref{Pz}(a) the axial velocities for electron $D_{e}$, ion $D_{i}$ and macroparticles $D_{d}$ fluids plotted when $B_{r0}=-1$ and $B_{z0}=1$, i.e. when the external axial magnetic field is permanent over the entire time interval of acceleration, and in Fig. \ref{Pz}(b) the same dependeces are presented for the case, when $B_{r0}=-1$ and $B_{z0}=-0,25t$, i.e. when external axial magnetic field is a constantly decreasing linear function of time.

\begin{figure}[H]
\begin{minipage}[h]{0.47\linewidth}
\begin{center}
\includegraphics[width=7cm]{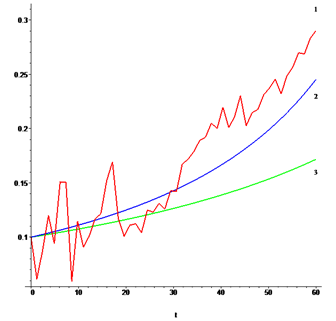} \\ (a)
\end{center}
\end{minipage}
\hfill
\begin{minipage}[h]{0.47\linewidth}
\begin{center}
\includegraphics[width=7cm]{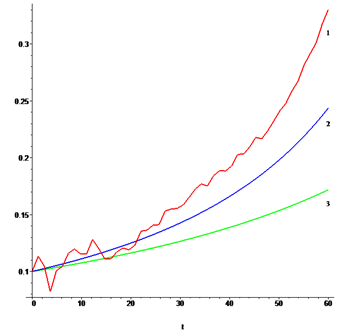} \\ (b)
\end{center}
\end{minipage}
\caption{Temporal dependence of $D_{e}$ (curve 1), $D_{i}$ (curve 2) and $D_{d}$ (curve 3) for $A_{e}(0)=A_{i}(0)=-10^{-2}$, $A_{d}=-7 \times 10^{-3}$, $-C_{e}(0)=-C_{d}=C_{i}(0)=10^{-3}$ for the case $B_{r0}=-1$ and $B_{z0}=1$ (a) and for the case $B_{r0}=-1$ and $B_{z0}=-0.25t$ (b). \label{Pz2}}
\end{figure}

Also, the dependence of the axial velocities of the dust plasma components on the magnitude of the radial velocity component was studied. In this case we took the following values of initial radial velocitiy components: $A_{e}(0)=A_{i}(0)=-10^{-2}$ and $A_{d}=-7 \times 10^{-3}$. The initial values of the azimuthal and axial velocity components were: $-C_{e}(0)=-C_{d}=C_{i}(0)=10^{-3}$ and $D_{e}(0)=D_{i}(0)=D_{d}=10^{-1}$. The initial values of the perturbed electric fields were $\varepsilon_{r}(0)=0$, $\varepsilon_{\phi}(0)=0$ and $\varepsilon_{z}(0)=0$, and the initial value of the perturbed axial magnetic field was $B^{*}_{z}(0)=0$. In Fig. \ref{Pz2}(a) the axial velocities for electron $D_{e}$, ion $D_{i}$ and macroparticles $D_{d}$ fluids plotted when $B_{r0}=-1$ and $B_{z0}=1$, and in Fig. \ref{Pz2}(b) the same dependecies are presented for the case, when $B_{r0}=-1$ and $B_{z0}=-0,25t$.

Namely, these dependencies illustrate what tipically occurs with the flow dynamics when the value $B_{z0}$ changes the charachter from permanent on a time function. As is seen from these graphs, in these configuration of magnetic fields each component of a dusty plasma is accelerated in axial direction. Here we observe the strong growth of axial velocities in the electron scale of time. In particular, it means that there is a weak influence of the collision phenomena on the momentum transfer if the character electron time is less than the corresponding characteristic collision time. Hence we can always find such plasma parameters and we may use the approximation of cold hydrodynamics for description of such processes. On the other hand, it can be noted that the radial velocity component exerts the strongest influence on the rate of axial acceleration of the macroscopic particles. This feature is required the further investigations.


\section{Conclusion}

In this paper, we considered the case of macroparticle acceleration for an already formed collisionless plasma flow containing electrons, positive singly charged ions and heavy multiply charged negative macroparticles in the accelerator based on nonequilibrium transfer of momentum between the macroscopic degrees of freedom of the plasma flow in crossed magnetic fields of a special type (see Fig. \ref{accelerator}(?)). A hydrodynamic model of the acceleration of a plasma flow containing electrons, protons and heavy multiply charged negative dust particles is proposed [see (\ref{dtn2})-(\ref{rotEcill2})]. This nonlinear model constructed on the basis of the equations of cold hydrodynamics using substitutions (\ref{v_s}), (\ref{E*}) and (\ref{n(t)}) allowed us to find the initial conditions under which the capture and acceleration of macroscopic particles takes place. These processes are caused by the action of external and intrinsic fields, which arise due to a local violation of quasineutrality [see (\ref{divEcill2})]. Thus, in this paper we propose a new scheme for the acceleration of macroscopic bodies and the principal possibility of accelerating dust particles is shown.

Here, we have restrict our consideration to the pure dynamical issue of the rotating plasma flow in the crossed magnetic fields. We don't discus the way to create the external magnetic fields of a desired form, we don't explore the process of plasma creation. However, even when the parameters are beyond the range of the model workability, the present conception of energy/momentum exchange may exist. Also, we would like to point out that the process of plasma production can be achieved by radial alternating electric field to propel a plasma discharge. However, in this case one should take into account the coupling of ionization processes and  nonlinear plasma oscillations \cite{ksch}. This may be used with a transformation of energy/momentum radial oscillations directing into the axial direction by the plasma inhomogeneity with nonlinear coupling among the electron and ion flow components and oscillations \cite{kys10}-\cite{Mirzaie}.

The present paper may be considered as a prerequisite to other more complete treatment for dynamics of rotating plasma flows in crossed non-stationary magnetic fields in the application to plasma thrusters.\\
The work of A.R.K and A.E.S. was supported by the Ministry of Science and Education of the Russian Federation under No. 14.575.21.0169 (RFMEFI57517X0169).

\end{document}